\documentclass{article}
\usepackage{t1enc}
\usepackage[utf8]{inputenc}
\usepackage[english]{babel}
\usepackage{graphicx}
\usepackage{amssymb}
\usepackage{amsmath}
\usepackage{amsthm}

\def\OA{{\cal A}}
\def\DK{{\cal O}}
\def\SDK{{\cal K}}
\def\PR{{\cal P}}
\def\vNA{{\cal N}}
\def\Mink{{\cal M}}
\def\HS{{\cal H}}
\def\CO{\mathbb{C}}
\def\RE{\mathbb{R}}
\def\IN{\mathbb{Z}}

\def\UN{\mathbf{1}}
\def\fel{{\frac{1}{2}}}

\def\qed{\ \vrule height 5pt width 5pt depth 0pt}
\def\cros{\raise1.9pt\hbox{$\scriptscriptstyle
          >$}\!\raise1.5pt\hbox{$\scriptstyle\triangleleft\,$}}
\theoremstyle{definition}\newtheorem{D}{Definition}
\theoremstyle{definition}
\theoremstyle{definition}\newtheorem{Prop}{Proposition}
\theoremstyle{definition}\newtheorem{Lemma}{Lemma}

\title{\bf Noncommutative Common Cause Principles \\ in Algebraic Quantum Field
Theory}
\author{\textit{G\'abor Hofer-Szab\'o}\thanks{King Sigismund College, Budapest,
email: gsz@szig.hu} \\
\textit{P\'eter Vecserny\'es}\thanks{Wigner Research Centre for Physics, Budapest, email: vecsernyes.peter@wigner.mta.hu}}
\date{ }

\begin{document}
\maketitle

\begin{abstract}
States in algebraic quantum field theory ''typically'' establish correlation
between spacelike separated events. Reichenbach's Common Cause Principle,
generalized to the quantum field theoretical setting, offers an apt tool to causally
account for these superluminal correlations. In the paper we motivate first
why commutativity between the common cause and the correlating events should be abandoned in the definition of the common cause. Then we show that the Noncommutative Weak Common Cause
Principle holds in algebraic quantum field theory with locally finite
degrees of freedom. Namely, for any pair of projections $A,B$ supported in
spacelike separated regions $V_A$ and $V_B$, respectively, there is a local
projection $C$ not necessarily commuting with $A$ and $B$ such that $C$ is
supported within the \textit{union} of the backward light cones of $V_A$ and $V_B$ and
the set $\{C,C^\perp\}$ screens off the correlation between $A$ and $B$.
\vspace{0.1in}

\end{abstract}

\section{Introduction} 

An operationally well motivated and mathematically transparent approach towards
quantum field theory is algebraic quantum field theory. In this theory
observables (including quantum events) are represented by $C^*$-algebras
associated to bounded regions of a given spacetime. The association of the
algebras and the spacetime regions is established along the following lines. 
\begin{enumerate}
\item \noindent\textit{Isotony.} Let $\mathcal{S}$ be a spacetime and let $\SDK$ be a collection of causally complete, bounded regions of $\mathcal{S}$, such that $(\SDK,\subseteq)$ is a directed poset under inclusion $\subseteq$. The net of local observables is given by the isotone map $\SDK\ni V\mapsto\OA(V)$ to unital $C^*$-algebras, that is $V_1 \subseteq V_2$ implies that $\OA(V_1)$ is a unital $C^*$-subalgebra of $\OA(V_2)$.

\item \noindent\textit{Quasilocal algebra.} The quasilocal observable algebra $\OA$ is defined to be the inductive limit $C^*$-algebra of the net $\{\OA(V),V\in\SDK\}$ of local $C^*$-algebras. 

\item \noindent\textit{Haag duality.} The net $\{\OA(V),V\in\SDK\}$ satisfies not only locality (Einstein causality) $\OA(V')'\cap\OA \supset \OA(V),V\in\SDK$ but also algebraic Haag duality: $\OA(V')'\cap\OA=\OA(V),V\in\SDK$, where primes denote spacelike complement and algebra commutant, respectively, and $\OA(V')$ is the smallest $C^*$-algebra in $\OA$ containing the local algebras $\OA(\tilde V), V'\supset\tilde V\in\SDK$.

\item \noindent\textit{Covariance.} Let $\mathcal{P}_\SDK$ be the subgroup of the group $\mathcal{P}$ of geometric symmetries of $\mathcal{S}$ leaving the collection $\SDK$ invariant. A group homomorphism $\alpha\colon\mathcal{P}_\SDK\to\textrm{Aut}\,\OA$ is given such that the automorphisms $\alpha_g,g\in\mathcal{P}_\SDK$ of $\OA$ act covariantly on the observable net: $\alpha_g(\OA(V))=\OA(g\cdot V), V\in\SDK$.
\end{enumerate}
To the net $\{\OA(V),V\in\SDK\}$ satisfying the above requirements we will refer
as a $\mathcal{P}_\SDK$-\textit{covariant local quantum theory}. If
$\mathcal{S}$ is a Minkowski spacetime $\Mink$ and $\SDK$ is the net of all
double cones in $\Mink$ then $\mathcal{P}_\SDK$ is the Poincar\'e group, and we obtain
Poincar\'e covariant algebraic quantum field theories with locally infinite
degrees of freedom. Restricting the collection $\SDK$ one can obtain
$\mathcal{P}_\SDK$-covariant local quantum theories with locally \textit{finite}
degrees of freedom (see Section 2). 

A \textit{state} $\phi$ in a local quantum theory is defined as a state
(normalized positive linear functional) on the quasilocal observable algebra
$\OA$. The corresponding GNS representation
$\pi_{\phi}\colon\OA\to\mathcal{B}(\mathcal{H})$ converts the net of
$C^*$-algebras into a net of subalgebras of $\mathcal{B}(\mathcal{H})$. Closing
these subalgebras in the weak topology one arrives at a net of local von Neumann
observable algebras: $\vNA(V):=\pi_{\phi}(\OA(V))'', V\in\SDK$. Here one can
require further properties such as unitary implementability of $\mathcal{P}_\SDK$ on $\mathcal{H}$, existence
of a vacuum representation, weak additivity, etc. (See Haag 1992). Since von
Neumann algebras are rich in projections, they offer a nice representation of
quantum \textit{events}: projections of a von Neumann algebra can be interpreted
as 0-1--valued observables where the expectation value of a projection is the
probability of the event that the observable takes on the value 1 in the given
quantum state. 

Two commuting events $A$ and $B$ (represented as projections) are said to be
\textit{correlating} in a state $\phi$ if $\phi(AB) \neq \phi(A)\phi(B)$. If the
events are supported in spatially separated spacetime regions $V_A$ and $V_B$,
respectively, then the correlation between them is said to be
\textit{superluminal}. A remarkable characteristics of Poincar\'e covariant
theories is that there exist ''many'' normal states establishing superluminal
correlations. (See (Summers, Werner 1988) or (Halvorson, Clifton 2000) for the
precise meaning of ''many''.) Since spacelike separation excludes direct causal
influence, one looks for a causal explanation of these superluminal correlations in terms of common causes.

Reichenbach (1956) characterizes the notion of the common cause in the following probabilistic way. Let $(\Sigma,p)$ be a classical probability measure space and
let $A$ and $B$ be two positively correlating events in $\Sigma$ that is let 
\begin{equation} \label{corr}
p(A\wedge B)>p(A)\, p(B) .
\end{equation}
\begin{D} \label{cc}
An event $C\in\Sigma$ is said to be the \textit{(Reichenbachian) common cause}
of the correlation between events $A$ and $B$ if the following  
conditions hold: 
\begin{eqnarray}
p(A\wedge B|C)&=&p(A|C)p(B|C) \label{cc1}\\
p(A\wedge B|C^{\perp})&=&p(A|C^{\perp})p(B|C^{\perp}) \label{cc2}\\
p(A|C)&>&p(A|C^{\perp})  \label{cc3}\\
p(B|C)&>&p(B|C^{\perp}) \label{cc4}
\end{eqnarray}
where $C^{\perp}$ denotes the orthocomplement of $C$ and $p( \, \cdot \,|\,
\cdot \,)$ is the conditional probability.
\end{D}
The above definition, however, is too specific since (i) it allows only for causes with a \textit{positive} impact on their effects and (ii) it excludes the possibility of a \textit{set} of cooperating common causes. Hence we need to generalize the definition of the common cause in the following manner:
\begin{D}\label{ccs}
A partition $\left\{ C_k \right\}_{k\in K}$ in $\Sigma$ is said to be the {\it
common cause system} of the correlation between events $A$ and $B$ if the
following screening-off condition holds for all $k\in K$: 
\begin{eqnarray} \label{ccs1}
p(A \wedge B\vert C_k)=p(A\vert C_k)\, p(B \vert C_k)
\end{eqnarray}
where the cardinality $|K|$ of $K$ is said to be the \textit{size} of the
common cause system. A common cause system of size $2$ is called a common cause
(without the adjective 'Reichenbachian', indicating that the inequalities
(\ref{cc3})-(\ref{cc4}) are not required).
\end{D}
As a next step, the notion of the common cause system is generalized to the
quantum case: First, one replaces the classical probability measure space
$(\Sigma,p)$ by the non-classical probability measure space $(\PR(\mathcal{N}),
\phi)$ where $\PR(\vNA)$ is the (non-distributive) lattice of projections
(events) and $\phi$ is a state of a von Neumann algebra $\vNA$. We note that in
case of projection lattices we will use only algebra operations (products,
linear combinations) instead of lattice operations ($\vee,\wedge$), because in
case of commuting projections $A,B\in\PR(\vNA)$ we have $A\wedge B=AB$ and
$A\vee B=A+B-AB$. 

A set of mutually orthogonal projections $\left\{ C_k \right\}_{k\in
K}\subset\mathcal{P}(\vNA)$ is called a \textit{partition of the unit}
$\UN\in\vNA$ if $\sum_k C_k = \UN$. Such a partition defines a
\textit{conditional expectation}
\begin{equation}\label{ncqcorr}
E\colon\vNA\to{\cal{C}}, \, \, A\mapsto 
  E(A):=\sum_{k\in K} C_kAC_k,
\end{equation}
that is $E$ is a unit preserving positive surjection onto the unital
$C^*$-subalgebra  ${\cal{C}}\subseteq\vNA$ obeying the bimodule property
$E(B_1AB_2)=B_1E(A)B_2; A\in\vNA, B_1, B_2\in{\cal{C}}$. We note that
${\cal{C}}$ contains exactly those elements of $\vNA$ that commute with
$C_k,k\in K$. Since $\phi\circ E$ is also a state on $\mathcal{N}$ we can give
the following  

\begin{D}\label{ncqccs}
A partition of the unit $\left\{ C_k \right\}_{k\in
K}\subset\mathcal{P}(\mathcal{N})$ is said to be a (possibly) {\em noncommuting common cause system} of the commuting events $A,B\in\cal{P}(\vNA)$, which correlate in the state
$\phi\colon\vNA\to\CO$, if 
\begin{eqnarray}\label{ncqccs1}
\frac{(\phi\circ E)(ABC_k)}{\phi(C_k)}&=& \frac{(\phi\circ E)(AC_k)}{\phi(C_k)}
\frac{(\phi\circ E)(BC_k)}{\phi(C_k)}. 
\end{eqnarray} 
for $k\in K$ with $\phi(C_k)\not= 0$. 
If $C_k$ commutes with both $A$ and $B$ for all $k\in K$ we call $\left\{ C_k
\right\}_{k\in K}$ a {\em commuting} common cause system. A common cause system of size $\vert K\vert=2$ is called
a {\em common cause}. 
\end{D}

Some remarks are in place here. First, in case of a commuting common cause
system $\phi\circ E$ can be replaced by $\phi$ in (\ref{ncqccs1}) since
$(\phi\circ E)(ABC_k)=\phi(ABC_k), k\in K$. Second, using the decompositions of
the unit, $\UN=A+A^{\perp}=B+B^{\perp}$, (\ref{ncqccs1}) can be rewritten in an
equivalent form:
\begin{equation}\label{ncqccsrew}
(\phi\circ E)(ABC_k))(\phi\circ E)(A^{\perp}B^{\perp}C_k)
=(\phi\circ E)(AB^{\perp}C_k)(\phi\circ E)(A^{\perp}BC_k),\ k\in K.
\end{equation}
One can even allow here the case $\phi(C_k)=0$ since then both sides of
(\ref{ncqccsrew}) are zero. Finally, it is obvious from (\ref{ncqccsrew}) that if $C_k\leq X$ with $X=A,A^\perp, B$ or $B^\perp$ for all $k\in K$ then $\left\{ C_k \right\}_{k\in K}$ serves as a (commuting) common cause system of the given correlation independently of the chosen state $\phi$. These solutions are called \textit{trivial common cause systems}. In case of common cause,
$\vert K\vert=2$, triviality means that $\{C_k\}=\{ A,A^\perp\}$ or $\{C_k\}=\{
B,B^\perp\}$. 

Attached to the definition of the common cause (system), Reichenbach's
Common Cause Principle (CCP) is the following hypothesis: \textit{if there is a
correlation between two events and there is no direct causal (or
logical) connection between the correlating events then there exists a common cause of the correlation.} The CCP in its present form, however, does \textit{not} refer to the spatiotemporal localization of the common cause (system). Since in a local quantum theory all
local events are supported in a well-defined spacetime region, the CCP
needs some tailoring to make it fit well to the local quantum field theoretical
setting. To address this point, one has to specify the localization of the
possible common causes of the correlations. One can define three different pasts
of the regions $V_1$ and $V_2$ in a spacetime $\mathcal{S}$ as (R\'edei, Summers 2007):
\begin{eqnarray*}\label{wcspast}
\makebox{{weak common past:}} &&  wpast(V_1, V_2) := I_-(V_1)\cup I_-(V_2) \\
\makebox{{common past:}} && cpast(V_1, V_2) := I_-(V_1)\cap I_-(V_2) \\
\makebox{{strong common past:}} && spast(V_1, V_2) := \cap_{x \in V_1 \cup
V_2}\, I_-(x),
\end{eqnarray*}
where $I_-(V)$ denotes the union of the backward light cones $I_-(x)$ of every
point $x$ in $V$ (R\'edei, Summers 2007). 
With these different localizations of the common cause in hand now we can define various
CCPs according to (i) whether commutativity is required and (ii) where the common cause system is localized. 
\begin{D}\label{CCP}
A $\mathcal{P}_\SDK$-covariant local quantum theory $\{\OA(V),V\in\SDK\}$ is
said to satisfy the Commu\-tative/Noncommutative (Weak/Strong) Common Cause
Principle if for any pair $A \in\mathcal A(V_1)$ and $B\in\mathcal A(V_2)$ of
projections supported in spacelike separated regions $V_1, V_2\in\SDK$ and
for every locally faithful state $\phi\colon\OA\to\CO$ such that
\begin{eqnarray}\label{qcorr'}
\phi(AB)  \neq  \phi(A)\, \phi(B)
\end{eqnarray}
there exists a \textit{nontrivial} commuting/noncommuting common cause system
$\{ C_k \}_{k\in K}\subset \mathcal A(V), V\in\SDK$ of the correlation
(\ref{qcorr'}) in the sense of Definition \ref{ncqccs} such that the
localization region $V$ is in the (weak/strong) common past of $V_1$ and $V_2$.
\end{D}

In the next Section we summarize the results concerning the Commutative CCPs. In
Section 3 we first motivate why commutativity is a physically unjustifiable
requirement for common cause systems in local quantum theories and why it should
be given up. Then our main theorem will be proven stating that the
\textit{Noncommutative Weak CCP} holds in algebraic quantum field theory with
locally finite degrees of freedom. Section 4 concludes the paper. 

\section{Commutative Common Cause Principles}

The question whether the Commutative Common Cause Principles are valid in a
Poincar\'e covariant local quantum theory in the von Neumann algebraic setting was first
raised by R\'edei (1997, 1998). As an answer to this question, R\'edei and
Summers (2002, 2007) have shown that the Commutative Weak CCP is valid in
algebraic quantum field theory with locally infinite degrees of freedom. Namely,
in the von Neumann setting they proved that for every locally normal and
faithful state and for every superluminally correlating pair of projections
there exists a weak common cause, that is a common cause system of size 2, in
the weak past of the correlating projections. They have also shown (R\'edei and
Summers, 2002, p 352) that the localization of the common cause cannot be
restricted to $wpast(V_1, V_2) \setminus I_-(V_1)$ or $wpast(V_1, V_2) \setminus
I_-(V_2)$. 

Concerning the Commutative (Strong) CCP less is known. If one also admits
projections localized only in \textit{un}bounded regions then the Strong CCP is
known to be false: von Neumann algebras pertaining to complementary wedges
contain correlated projections but the strong past of such wedges is empty (see
Summers and Werner, 1988 and Summers, 1990). However, restricting ourselves to
\textit{local} algebras the situation is not clear. We are of the opinion that one
cannot decide on the validity of the (Strong) CCP without an explicit reference
to the dynamics since there is no bounded region $V$ in $cpast(V_1, V_2)$ (hence
neither in $spast(V_1, V_2)$) for which isotony would ensure that $\OA(V_1\cup
V_2)\subset\OA(V'')$. But dynamics relates the local algebras since $\OA(V_1\cup
V_2)\subset\OA(V''+t) =\alpha_t(\OA(V''))$ can be fulfilled for certain $V\in
cpast(V_1, V_2)$ and for certain time translation by $t$. 

Coming back to the proof of R\'edei and Summers, the proof had a crucial premise,
namely that the algebras in question are \textit{von Neumann algebras of type
III}. Although these algebras arise in a natural way in the context of
Poincar\'e covariant theories other local quantum theories apply von Neumann
algebras of other type. For example, theories with locally finite degrees of
freedom are based on finite dimensional (type I) local von Neumann algebras.
This raised the question whether the Commutative Weak CCP is valid in other
local quantum theories. To address the problem Hofer-Szab\'o and Vecserny\'es (2012) have chosen the local quantum Ising model (see M\"uller, Vecserny\'es) having locally finite degrees of freedom. It turned out that the Commutative Weak CCP is \textit{not valid} in the local quantum Ising model and it cannot be valid either in theories with locally finite degrees of freedom in general.

\begin{figure}[ht]
\centerline{\resizebox{6cm}{!}{\includegraphics{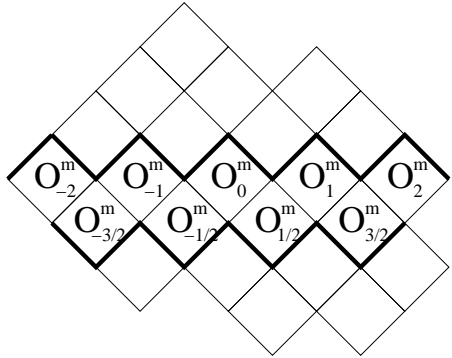}}}
\caption{A thickened Cauchy surface in the two dimensional Minkowski space
$\Mink^2$}
\label{cauchyV}
\end{figure}

Since this model, which is a prototype of local UHF-type quantum
theories, will play a role in our theorem proven in the next Section, we
introduce it here. First, consider the net of `intervals'
$(i,j):=\{i,i+\frac{1}{2},\dots,j-\frac{1}{2},j\}\subset\frac{1}{2}\mathbb{Z}$
of half-integers. The set of half-integers can be interpreted as the space
coordinates of the center $(0,x),x\in\mathbb{Z}$ and
$(-1/2,x),x\in\mathbb{Z}+1/2$ of minimal double cones $\DK^m_x$ of unit diameter
on a 'thickened' Cauchy surface in two dimensional Minkowski space $M^2$ (See
Fig. \ref{cauchyV}.) An interval $(i,j)\subset\frac{1}{2}\mathbb{Z}$ can be
interpreted as the smallest double cone $\DK_{i,j}\subset M^2$ containing both
$\DK_i^m$ and $\DK_j^m$. They determine a directed subset $\SDK^m_{CS}$ of
double cones in $\Mink^2$, which is left invariant by the group of
space-translations with integer values. 

The net of local algebras is defined as follows. Consider the `one-point'
observable algebras $\OA(\DK^m_i), \, \, i\in\frac{1}{2}\mathbb{Z}$ associated
to the minimal double cone $\DK^m_i$. Let $U_i$ denote the selfadjoint unitary
generator of the algebra $\OA(\DK^m_i)\simeq M_1(\CO)\oplus M_1(\CO)$ for any
$i\in\frac{1}{2}\mathbb{Z}$. Between the generators one demands the following
commutation relations:
\begin{eqnarray}\label{comm_rel}
U_i U_j = \left\{ \begin{array}{rl} -U_j U_i, & \mbox{if}\ |i-j|=\frac{1}{2},
\\ U_j U_i, & \mbox{otherwise.}\ \end{array} \right.
\end{eqnarray}
Now, the local algebra $\OA(\DK_{i,j})$ associated to the double cone
$\DK_{i,j}\in\SDK^m_{CS}$ are linearly spanned by the monoms
\begin{eqnarray}\label{monoms}
U_i^{k_i} \, U_{i+\frac{1}{2}}^{k_{i+\frac{1}{2}}} \,  \dots \,
U_{j-\frac{1}{2}}^{k_{j-\frac{1}{2}}} \, U_j^{k_j} 
\end{eqnarray}
where $k_i, k_{i+\frac{1}{2}} \dots k_{j-\frac{1}{2}}, k_j \in \{0,1\}$.  Since
the local algebras $\OA(\DK_{i,i-\frac{1}{2}+n}), i\in\frac{1}{2}\mathbb{Z}$ for
$n\in \mathbb{N}$ are isomorphic to the full matrix algebra
$M_{2^n}(\mathbb{C})$ the quasilocal observable algebra $\OA$ is a uniformly
hyperfinite (UHF) $C^*$-algebra.\footnote{For detailed Hopf algebraic
description of the local quantum spin models see (Szlach\'anyi, Vecserny\'es,
1993), (Nill, Szlach\'anyi, 1997), (M\"uller, Vecserny\'es))}

In case of the Ising model the causal (integer valued) time evolutions are
classified (see M\"uller, Vecserny\'es). That is the possible causal and
(discrete) time translation covariant extensions of the ''Cauchy surface net'' 
$\{\OA(\DK),\DK\in\SDK^m_{CS}\}$ to $\{\OA(\DK),\DK\in\SDK^m\}$ are given, where
$\SDK^m$ is the subset of double cones in $\Mink^2$ that are spanned by minimal
double cones being integer time translates of those in $\SDK^m_{CS}$, i.e. those
in the original Cauchy surface. The set $\SDK^m$ is left invariant by integer
space and time translations, and the extended net also satisfies isotony,
Einstein causality and algebraic Haag duality: $\OA(\DK')'\cap\OA
=\OA(\DK),\DK\in\SDK^m$. Moreover, the commuting (unit) time and (unit) space
translation automorphisms $\beta$ and $\alpha$ of the quasilocal algebra $\OA$
act covariantly on the local algebras. The causal time translation automorphisms
$\beta$ of $\OA$ can be parametrized by $\theta_1,\theta_2;\eta_1,\eta_2$ with
$-\pi/2 <\theta_1,\theta_2\leq\pi/2$ and $\eta_1,\eta_2\in\{1,-1\}$ and they are
given on the algebraic generator set $\{U_i\in\OA(\DK^m_i),
i\in\frac{1}{2}\mathbb{Z}\}$ of $\mathcal A$. The  automorphisms
$\beta=\beta(\theta_1,\theta_2,\eta_1,\eta_2)$ of $\mathcal A$ corresponding to
causal time translations by a unit read as
\begin{eqnarray}\label{causal_automorph1}
\beta(U_x)&=&\eta_1 \sin^2\theta_1 U_x+\eta_1 
      \cos^2\theta_1 U_{x-\fel}U_xU_{x+\fel}\nonumber\\
  &&\quad+\frac{i}{2}\sin2\theta_1(U_{x-\fel}U_x-U_xU_{x+\fel}),\\
\label{causal_automorph2}
\beta(U_{x+\fel})&=&\eta_2\sin^2\theta_2 U_{x+\fel} 
         +\eta_2\cos^2\theta_2\beta(U_x)U_{x+\fel}
         \beta(U_{x+1})\nonumber\\
 &&\quad+\frac{i}{2}\sin2\theta_2(\beta(U_x)U_{x+\fel}-
  U_{x+\fel}\beta(U_{x+1})),
\end{eqnarray}
where $x\in\IN$. The causal evolutions clearly show how local primitive
causality holds in the discrete case: If $V$ consists of three neighbouring
minimal double cones on a thickened Cauchy surface then $\OA(V)=\OA(V'')$ due to
(\ref{causal_automorph1}) and (\ref{causal_automorph2}). Moreover, the following
algebra isomorphisms hold (see M\"uller, Vecserny\'es): If $\DK\in\SDK^m$ is a
double cone containing $n_+$ and $n_-$ minimal double cones in the right forward
and left forward lightlike directions, respectively, then $\vert\OA(\DK)\vert$,
the linear dimension of the corresponding local algebra is $2^{n(\DK)},
n(\DK):=n_++n_--1$ and
\begin{eqnarray}\label{localgtype}
\OA(\DK) \simeq \left\{ \begin{array}{rl} 
M_{2^{n(\DK)/2}}(\CO), & \mbox{if}\ n(\DK)\ \mbox{is even,}\\ 
M_{2^{(n(\DK)-1)/2}}(\CO)\oplus M_{2^{(n(\DK)-1)/2}}(\CO), 
                       & \mbox{if}\ n(\DK)\ \mbox{is odd.} \end{array} \right.
\end{eqnarray}
These properties lead us to the following definition:
\begin{D}\label{localUHFtype} Let $W^{L/R}(\DK)$ and $I_\pm(\DK)$ be the
smallest left/right wedge region and the smallest forward/backward light cone, respectively, 
in the 2-dimensional Minkowski space $\Mink^2$ that contain the double cone
$\DK\in\SDK^m$. The local quantum theory $\{\OA(\DK),\DK\in\SDK^m\}$ in
$\Mink^2$ is called \textit{local UHF-type quantum} theory if 
\begin{itemize}
\item[(i)] $\OA(\DK), \DK\in\SDK^m$ are finite dimensional $C^*$-algebras, 
\item[(ii)] for any $\DK\in\SDK^m$ there exist double cones
$\DK^{L/R}_\pm\in\SDK^m$ with $\DK\subset\DK^{L/R}_\pm\subset W^{L/R}(\DK)\cap
I_\pm(\DK)$ such that $\OA(\DK^{L/R}_\pm)$ are simple $C^*$-algebras,
\item[(iii)] local primitive causality holds: if $V$ consists of three
neighbouring minimal double cones on a thickened Cauchy surface then
$\OA(V)=\OA(V'')$.
\end{itemize}
\end{D}
Clearly, the quasilocal algebra $\OA$ of a local UHF-type quantum theory is a
UHF $C^*$-algebra. What is more, any two local algebras with spacelike
separated supports $\DK_1, \DK_2\in\SDK^m$ are embedded in simple local algebras
with spacelike separated supports contained in the weak future/past
$I_\pm(\DK_1)\cup I_\pm(\DK_2)$ of $\DK_1$ and $\DK_2$. Local primitive
causality implies that for any triple $\DK_1,\DK_2,\tilde\DK_\pm\in\SDK^m$ with 
$\DK_1\vee\DK_2\subset\tilde\DK_\pm\subset I_\pm(\DK_1\vee\DK_2)$ the algebras obey $\OA(\tilde\DK_\pm) = \OA(\tilde\DK_\pm\cap I_\pm(\DK_1))\vee\OA(\tilde\DK_\pm\cap I_\pm(\DK_2)) $.

Based on this $\mathbb Z \times \mathbb Z$-covariant local quantum Ising model Hofer-Szab\'o and Vecserny\'es (2012) proved the following Proposition:
\begin{Prop}\label{failWCCP}
A local UHF-type quantum theory $\{\OA(\DK),\DK\in\SDK^m\}$ in $\Mink^2$ does
\textit{not} satisfy the Commutative Weak CCP, in general. That is there exist nonzero
projections $A\in\OA(\DK_a)$ and $B\in\OA(\DK_b)$ localized in two spacelike
separated double cones $\DK_a,\DK_b\in\SDK^m$ and a faithful state on $\OA$
establishing a correlation between $A$ and $B$ such that there exists no
nontrivial common cause system of the correlation localized in $wpast(\DK_a,
\DK_b)$.
\end{Prop}
Hence, in contrast with local quantum theories having locally \textit{infinite}
degrees of freedom, the Commutative Weak CCP is \textit{not} valid in general in
models having locally \textit{finite} degrees of freedom. Obviously, the Weak
CCP is weaker then the (Strong) CCP, hence the Proposition above falsified the
other two Commutative CCPs as well. 

As it was argued at the end of (Hofer-Szab\'o and Vecserny\'es 2012), one can
react to this fact in two different ways. On the one hand, one may regard a
discrete model as an approximation of a ''more detailed'' (discrete or
continuous) model. Then the failure of Commutative CCPs is explained by the fact
that the approximate model does not contain all the observables; the common
cause remains burried beyond the coarse description of the physical situation in
question. Only a refined extended model could reveal the hidden common causes. 

The other strategy, however, is to accept the discrete model as a self-contained physical model describing a specific physical phenomenon\footnote{See e.g. (Borthwick, Garibaldi, 2010) as a recent exprimental development concerning the Ising model.} and to endorse the consequence of Proposition \ref{failWCCP} as to the Commutative CCP cannot be a universally valid principle in local quantum theories. But this immediately raises the following question: What is the situation if we abandon commutativity between the common cause (system) and the correlating events? Are the Noncommutative CCPs valid in AQFT?

In (Hofer-Szab\'o and Vecserny\'es 2012) there have been some indications that replacing commuting common causes by noncommuting ones can help to regain the CCP: for the same state falsifying the Commutative Weak CCP a noncommuting common cause could be given (located in the common past of the correlating events). In the next Section we will show that this specific example is indeed the sign of a more general fact, namely the validity of the Noncommutative Weak CCP.

\section{Noncommutative Common Cause Principles}

Commutativity has a well-specified role in quantum theories. Observables should
commute to be simultaneously measurable in quantum mechanics; commutativity of
observables with spacelike separated supports is one of the axioms of local
quantum theories. However, as the possible causal time evolutions
(\ref{causal_automorph1})-(\ref{causal_automorph2}) in the local quantum Ising
model show, the local observable $U_x$ does not commute even with its
own unit time translates $\beta(U_x)$ (unless $\theta_1=0$ or $\pi/2$). This
happens already in ordinary quantum mechanics.

Let us consider the quantum
harmonic oscillator given by the 
Hamiltonian $H$
\begin{equation}\label{oscHam}
H=-\frac{\hbar}{2m}\frac{d^2}{dx^2} +\frac{1}{2} m\omega^2 x^2
\end{equation}
acting on the Hilbert space $\cal H$ of square integrable functions on the real
line $\RE$. The eigenfunctions $\psi_n\in\HS, n=0,1,2,\dots$ of $H$ are 
\begin{equation}\label{efuncHam}
\psi_n(x)=\sqrt{\frac{1}{2^n n!}}
  \left(\frac{m\omega}{\pi\hbar}\right)^{\frac{1}{4}}
  \exp{\left(\frac{-m\omega x^2}{2\hbar}\right)}
  H_n\left( \sqrt{\frac{m\omega}{\hbar}}x\right), 
\end{equation}
where $H_n$ is the Hermite polynomial
\begin{equation}
H_n(x)=(-1)^n\exp{\left({x^2}\frac{d^n}{dx^n}\right)} \exp{\left(-x^2\right)}.
\end{equation}
The corresponding energy eigenvalues are $E_n=\hbar\omega(n+1/2)$.
The unitary time evolution operator $U(t):=\exp{(iHt)}$ is diagonal in the basis
$\{\psi_n\}\subset\HS$ with entries $\exp{(iE_nt)}$.
Using the recursion relation
\begin{equation}
xH_n(x)=\frac{1}{2}H_{n+1}(x)+nH_{n-1}(x)
\end{equation}
for the Hermite polynomials, 
it is easy to see that the time evolved position operator 
$x(t):=U(-t)xU(t)$ does not commute with $x\equiv x(0)$ for generic $t$,
i.e for $\hbar\omega t\not\in 2\pi\IN$. For example, in the ground state
\begin{eqnarray}
\frac{m\omega}{\hbar}\, \big[ x, x(t)\big] \, \psi_0
 &=&(e^{i(E_0-E_1)t}-e^{i(E_1-E_2)t})\frac{1}{\sqrt{2}}\psi_2
    +(e^{i(E_0-E_1)t}-e^{i(E_1-E_0)t})\frac{1}{2}\psi_0\nonumber\\
 &=&-i\sin{(\hbar\omega t)}\psi_0 \not\equiv 0.
\end{eqnarray}
Thus, if an observable $A$ is not a conserved quantity, that is $A(t)\not= A$,
then the commutator $[A,A(t)]\not= 0$ in general. So why should the commutators
$[A,C]$ and $[B,C]$ vanish for the events $A,B$ and for their common cause
$C$ supported in their (weak/common/strong) past? We think that commuting common causes are only unnecessary reminiscense of their classical formulation. Due to their relative spacetime localization, that is due to the time delay between the correlating events and the common cause, it is also an unreasonable assumption. The benefit of allowing
noncommuting common causes is that they help to maintain the validity of the Weak
CCP also in local UHF-type quantum theories.

\begin{Lemma}\label{ncCC}
Let $\vNA$ be a tensor product matrix algebra $\vNA=M_{n_1}(\CO)\otimes
M_{n_2}(\CO)$. Let $\vNA_1$ and $\vNA_2=\vNA_1'\cap\vNA$ denote the unital
subalgebras $M_{n_1}(\CO)\otimes\UN$ and $\UN\otimes M_{n_2}(\CO)$,
respectively. Let $\phi$ be a faithful state on $\vNA$ that leads to a
correlation between two nontrivial (commuting) projections $A\in\vNA_1$ and
$B\in\vNA_2$. Then there exists a (noncommuting) common cause $\{ C_1\equiv C,
C_2\equiv C^\perp\}\subset{\cal{P}}(\vNA)$ such that
\begin{equation}\label{ncqccsrew1}
\phi(C_kABC_k)\, \phi(C_k A^{\perp}B^{\perp}C_k)
=\phi(C_kAB^{\perp}C_k)\,\phi(C_kA^{\perp}BC_k),\ k=1,2.
\end{equation}
\end{Lemma}

\noindent\textit{Proof.} Let $Tr$ be the (unique) normalized trace on $\vNA$. We
can suppose that $Tr B\geq Tr B^\perp$. (If it fails then  change the pair $A,
B$ to $A^\perp, B^\perp$, (\ref{ncqccsrew1}) is invariant with respect to this
change.) Let us choose the projection $C:=AB'$ such that $B'\in\vNA_2$ is an
equivalent projection to $B$, $B'\sim B$, that is $Tr B'= Tr B$. Clearly $C$ is
a subprojection of $A$, $C<A$, therefore (\ref{ncqccsrew1}) fulfills for $k=1$
because both sides are zero. 
For $k=2$ one obtains
\begin{equation}\label{Cperp1}
\phi(C^\perp ABC^\perp)\phi(A^\perp B^\perp)
=\phi(C^\perp AB^\perp C^\perp)\phi(A^\perp B)
\end{equation}
since $C^\perp > A^\perp$. Substituting $B=\UN-B^\perp$ into (\ref{Cperp1}) one
arrives at
\begin{equation}\label{Cperp2}
\frac{\phi(A^\perp B^\perp)}{\phi(A^\perp)}
=\frac{\phi(C^\perp AB^\perp C^\perp)}{\phi(C^\perp A)}.
\end{equation}
Due to faithfulness of $\phi$ the left hand side is in the interval $(0,1)$. Let
us choose a continuous map $U\colon [0,1]\to U(\vNA_2)$ from the unit interval
into the unitary elements of $\vNA_2$ such that $U(0)=\UN$ and
$B(1):=U(1)BU(1)^*\geq B^\perp$. Since projections having the same dimensions
(the same trace) in a simple algebra $\vNA_2$ are equivalent and the group of
unitaries in $\vNA_2$ is connected, such a path exists. Then 
\begin{equation}\label{Cpath}
[0,1]\ni t\mapsto C(t):=AU(t)BU(t)^*=U(t)ABU(t)^*\in{\cal{P}}(\vNA)
\end{equation}
defines a continuous map with $C(0)=AB$ and $C(1)=AB(1)\geq AB^\perp$. Hence, 
\begin{equation}\label{fracCpath}
[0,1]\ni t\mapsto F(t):=\frac{\phi(C(t)^\perp AB^\perp
C(t)^\perp)}{\phi(C(t)^\perp A)}\in [0,1]
\end{equation}
defines a continuous function with $F(0)=1$ and $F(1)=0$ since $C(0)^\perp
A=AB^\perp$ (hence, in the nominator $C(0)^\perp AB^\perp C(0)^\perp=AB^\perp$)
and $C(1)^\perp A\leq AB$ (hence, in the nominator $C(1)^\perp AB^\perp
C(1)^\perp=0$), respectively. Therefore there exists $t'\in (0,1)$ such that
$F(t')$ is equal to the left hand side of (\ref{Cperp2}) and the corresponding
pair of projections $\{ AB':=C(t'), C(t')^\perp\}$ serves as a common cause.\qed

The lemma above is enough to prove the validity of the Noncommutative Weak CCP in the local
quantum Ising model for all possible causal time evolutions. We use only that
they are local UHF-type quantum theories for all causal time evolutions.
\begin{Prop}\label{ncWCCP}
Let $\{\OA(\DK),\DK\in\SDK^m\}$ be a local UHF-type quantum theory in $\Mink^2$.
Let  $A\in\OA(\DK_a)$ and $B\in\OA(\DK_b)$ be projections with spacelike
separated supports $\DK_a,\DK_b\in\SDK^m$. If $\phi$ is a locally
faithful state on $\OA$ leading to a correlation between $A$ and $B$ then there
exists a common cause $\{ C,C^\perp\}$ localized in the weak past of $\DK_a$ and
$\DK_b$. 
\end{Prop}

\noindent\textit{Proof.} We can suppose $\DK_a$ to be on the left of $\DK_b$,
that is $\DK_a\in W^L(\DK_b)$. Since we are dealing with a local UHF-type
quantum theory there exist spacelike separated double cones $\DK_{a-}^L,\DK_{b-}^R\in\SDK^m$ with $\DK_a\subset\DK_{a-}^L\subset
I_-(\DK_a)\cap W^L(\DK_a)$ and $\DK_b\subset\DK_{b-}^R\subset I_-(\DK_b)\cap W^R(\DK_b)$
with simple local algebras $\OA(\DK_{a-}^L)$ and $\OA(\DK_{b-}^R)$, respectively. 
Moreover, if
$\tilde\DK:=\DK_{a-}^L\vee\DK_{b-}^R\in\SDK^m$ then there exists
$\tilde\DK_-^R\in\SDK^m$ with $\tilde\DK\subset\tilde\DK_-^R\subset I_-(\DK_a\vee\DK_b)$ with simple local algebra $\OA(\tilde\DK_-^R)$, which is equal to 
$\OA(\tilde\DK_-^R\cap I_-(\DK_a))\vee \OA(\tilde\DK_-^R\cap I_-(\DK_b))$ due to local primitive causality. 

Defining the simple algebras $\vNA_1:=\OA(\DK_{a-}^L)$ and
$\vNA:=\OA(\tilde\DK_-^R)$, the algebra $\vNA_2:=\vNA_1'\cap\vNA$, being a
commutant of a simple unital subalgebra in a simple matrix algebra, is simple
and contains $\OA(\DK_{b-}^R)$. Since the restriction of the state $\phi$ to
$\vNA$ has been required to be faithful one can apply Lemma \ref{ncCC} for the
quantities $\vNA, A\in\vNA_1, B\in\vNA_2$ to prove the existence of a common
cause $\{ C,C^\perp\}$ in $\OA(\tilde\DK_-^R)$, that is in the weak past of
$\DK_a$ and $\DK_b$.\qed

A direct consequence of Proposition \ref{ncWCCP} is that the Noncommutative Weak CCP holds in local UHF-type quantum theories. It is obvious from the proof that the given common cause $\{ C,C^\perp\}$ in ${wpast}(\DK_a,\DK_b)$ is not sensitive to the explicit dynamics: we have used only local primitive causality of the dynamics and isotony of the local algebras. Namely, using the notations of the previous proof, both the correlating events and the common cause are elements of the algebra $\OA(\tilde\DK_-^R)=\OA(\tilde\DK_-^R\cap I_-(\DK_a))\vee \OA(\tilde\DK_-^R\cap I_-(\DK_b))$. This property is missing in the case of the CCP, thus we discuss the consequences shortly in the rest of this Section.

\begin{figure}[ht]
\centerline{\resizebox{6.3cm}{!}{\includegraphics{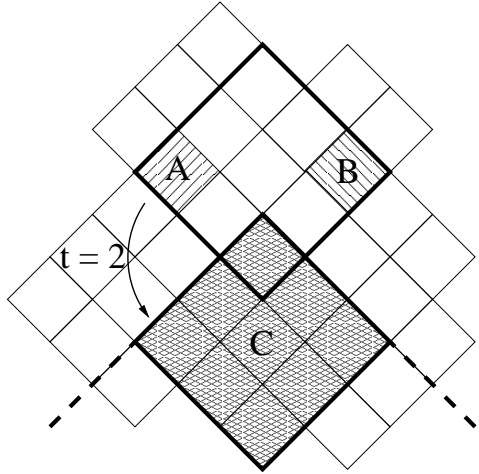}}}
\caption{A plausible localization of a common cause in the common past of the correlating events $A$ and $B$.}
\label{strong_Bell1111}
\end{figure}

The reason why the other CCPs are a more subtle problem than the weak CCP is that the algebra $\OA(\DK_a)\vee\OA(\DK_b)$ is not contained in any local algebra supported in ${cpast}(\DK_a,\DK_b)$. Isotony does not help, because $\DK_a\cup\DK_b\not\subset {cpast}(\DK_a,\DK_b)$. Hence, we need the explicit dynamics to relate the correlating events in $\OA$ to a common cause supported in ${cpast}(\DK_a,\DK_b)$. Although it is not clear how far one has to go back in the common past ${cpast}(\DK_a,\DK_b)$ for finding a common cause (if it exists at all) located in $\DK_c$ as a first guess one can use $\DK_c:=\DK_a\vee\DK_b-(t,0)\in\SDK^m$, i.e. the time shifted double cone generated by $\DK_a$ and $\DK_b$, for the smallest time $t$ for which $\DK_c\subset{cpast}(\DK_a,\DK_b)$. (See Fig. \ref{strong_Bell1111}.)
Since a proof of the existence or non-existence of a common causes in ${cpast}(\DK_a,\DK_b)$ seems to be difficult some examples may give a hint. 

\begin{figure}[ht]
\centerline{\resizebox{6cm}{!}{\includegraphics{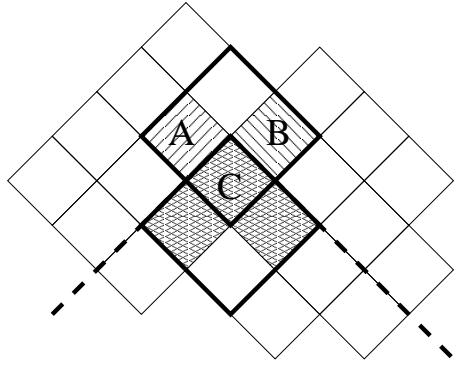}}}
\caption{Localization of the noncommuting common cause of the correlation $(A,B)$ for a special dynamics.}
\label{strong_Bell11}
\end{figure}

In our previous paper (Hofer-Szab\'o, Vecserny\'es 2012) we have found common causes for events $A,B$ localized in two adjacent but spacelike separated minimal double cones $\DK_a=\DK^m_{-\fel} + (1,0)$ and $\DK_b=\DK^m_{\fel} + (1,0)$ for one of the simplest dynamics of (\ref{causal_automorph1})-(\ref{causal_automorph2}), namely for $\theta_1=0,\eta_1=1$. (See Fig. \ref{strong_Bell11}.)
The faithful states $\phi_\lambda(\, \cdot \,):=Tr(\rho_\lambda\, \cdot \,)$ on $\OA$ parametrized by $\lambda=(\lambda_1,\lambda_2,\lambda_3,\lambda_4)$, which lead to correlation between $A$ and $B$, were given by the density matrices 
\begin{equation}\label{corrstates}
\rho_\lambda=\lambda_1 AB+\lambda_2 A^\perp B^\perp+\lambda_3 A^\perp B+\lambda_4 AB^\perp, \quad\lambda_i> 0,\ \sum_{i=1}^4\lambda_i=4,\ \lambda_1\lambda_2 \in \mathbb{Q},\ \lambda_3\lambda_4 \notin \mathbb{Q}
\end{equation}
with the restriction $\lambda_1+\lambda_2=\lambda_3+\lambda_4$. The common cause was
supported in $\DK_a\vee\DK_b-(1,0)\subset {cpast}(\DK_a,\DK_b)$ fulfilling the first guess mentioned above. However, for generic time evolutions (\ref{causal_automorph1})-(\ref{causal_automorph2}) we have not found a common cause localized in this region for all the states given in (\ref{corrstates}).

\begin{figure}[ht]
\centerline{\resizebox{6cm}{!}{\includegraphics{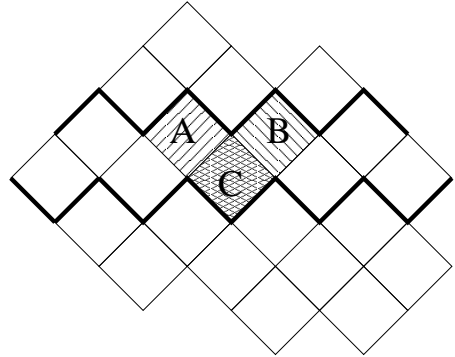}}}
\caption{Localization of the noncommuting common cause of the correlation $(A,B)$ for all dynamics.}
\label{strong_Bell111}
\end{figure}

Requiring also $\lambda_1=\lambda_2$ the set $\{ C,C^\perp\}=\{\fel(\UN\pm U_0)\}$ localized in $\DK^m(0,0)\subset \DK_a\vee\DK_b-(1,0)$ serves already as a common cause for \textit{all} dynamics. (See Fig. \ref{strong_Bell111}.)
But the shortcoming of this solution is that the common cause and the correlating events lie within a common Cauchy surface, i.e. the solution is not sensitive to the dynamics.

\section{Conclusions}

In the paper we have shown that the Weak Common Cause Principle can be maintained also in local quantum theories with locally finite degrees of freedom \textit{if} one allows noncommuting common causes as well. Since observables do \textit{not} commute even with its own time translates in general and since the common causes are required to supported in the (weak) past of the correlating events we think that noncommuting generalization of common causes is a natural assumption in quantum theories. The investigation of the (Strong) Common Cause Principles needs some knowledge about the dynamics, hence their validity is a more difficult problem. Maybe in the simplest local quantum theory, in the local quantum Ising model where the possible dynamics are known one can tackle this problem as well.

Finally, we mention a further possible direction of research. As we saw in this paper, abandoning commutativity gives us extra freedom in the search of common causes for correlations. But how big is this freedom? Is it big enough to find a common cause for a \textit{set} of correlations? Or in other words, if $\{(A_m, B_n) \, \vert \, m\in M, n\in N\}$ is a set of correlating events supported in spacelike separated regions $V_A$ and $V_B$, respectively, then does there exist a (weak/strong) \textit{common} common cause system of these correlations, that is a common cause system screening off \textit{all} the correlations? 
\vspace{0.2in}

\noindent
{\bf Acknowledgements.} This work has been supported by the Hungarian Scientific Research Fund, OTKA K-68195 and by the Fulbright Research Grant while G. H-Sz. was a Visiting Fellow at the Center for Philosophy of Science in the University of Pittsburgh.

\section*{References} 
\begin{list}
{ }{\setlength{\itemindent}{-15pt}
\setlength{\leftmargin}{15pt}}

\item D. Borthwick and S. Garibaldi, ''Did a 1-dimensional magnet detect
248-dimensional Lie algebra?'' \textit{Not. Amer. Math. Soc.} \textbf{58}, 1055-1066 (2011) 

\item R. Haag, {\it Local Quantum Physics}, (Springer Verlag, Berlin, 1992). 

\item H. Halvorson and R. Clifton, ''Generic Bell correlation between arbitrary
local algebras in quantum field theory,''
\textit{J. Math. Phys.}, \textbf{41}, 1711-1717 (2000).

\item G. Hofer-Szab\'o and P. Vecserny\'es, ''Reichenbach's Common Cause Principle in
algebraic quantum field theory with locally finite degrees of freedom,''
\textit{Found. Phys.}, \textbf{42}, 241-255 (2012).

\item V.F. M\"uller and P. Vecserny\'es, ''The phase structure of $G$-spin
models'', \textit{to be published}

\item F. Nill and K. Szlach\'anyi, ''Quantum chains of Hopf algebras with
quantum double cosymmetry'' \textit{Commun. Math. Phys.}, \textbf{187} 159-200
(1997).

\item M. R\'edei, ''Reichenbach's Common Cause Principle and quantum field
theory,'' \textit{Found. Phys.}, \textbf{27}, 1309--1321 (1997).

\item M. R\'edei, {\it Quantum Logic in Algebraic Approach}, (Kluwer Academic
Publishers, Dordrecht, 1998).

\item M. R\'edei and J. S. Summers, ''Local primitive causality and the Common
Cause Principle in quantum field theory,'' \textit{Found. Phys.}, \textbf{32},
335-355 (2002).

\item M. R\'edei and J. S. Summers, ''Remarks on Causality in relativistic
quantum field theory,'' \textit{Int. J. Theor. Phys.}, \textbf{46}, 2053–2062
(2007).

\item H. Reichenbach, {\it The Direction of Time}, (University of California
Press, Los Angeles, 1956).

\item S. J. Summers, ''On the independence of local algebras in quantum field
theory,'' \textit{Reviews in Mathematical Physics}, \textbf{2}, 201-247 (1990).

\item S. J. Summers and R. Werner, ''Maximal violation of Bell's inequalities
for algebras of observables in tangent spacetime regions,'' \textit{Ann. Inst.
Henri Poincar\'e -- Phys. Th\'eor.}, \textbf{49}, 215-243 (1988).

\item K. Szlach\'anyi and P. Vecserny\'es, ''Quantum symmetry and braid group
statistics in $G$-spin models'' \textit{Commun. Math. Phys.}, \textbf{156},
127-168 (1993). 

\end{list}

\end{document}